# The JuliaConnectoR: A Functionally Oriented Interface for Integrating **Julia** in R

Stefan Lenz[1*], Maren Hackenberg[1], Harald Binder[1]

[1]Institute of Medical Biometry and Statistics,
Faculty of Medicine and Medical Center – University of Freiburg
[*]Corresponding author: lenz@imbi.uni-freiburg.de

## Abstract

Like many groups considering the new programming language Julia, we faced the challenge of accessing the algorithms that we develop in Julia from R. Therefore, we developed the R package **JuliaConnectoR**, available from the official R package repository and from GitHub (https://github.com/stefan-m-lenz/JuliaConnectoR), in particular for making advanced deep learning tools available. For maintainability and stability, we decided to base communication between R and Julia via the transmission control protocol, using an optimized binary format for exchanging data. Our package also specifically contains features that allow for a convenient interactive use in R. This makes it easy to develop R extensions with Julia or to simply call functionality from Julia packages in R. Interacting with Julia objects and calling Julia functions becomes user-friendly, as Julia functions and variables are made directly available as objects in the R workspace. We illustrate the further features of our package with code examples, and also discuss advantages over the two alternative packages **JuliaCall** and **XRJulia**. Finally, we demonstrate the usage of the package with a more extensive example for employing neural ordinary differential equations, a recent deep learning technique that has received much attention. This example also provies more general guidance for integrating deep learning techniques from Julia into R.

## 1. Introduction

R (R Core Team 2020a) and Julia (Bezanson, Edelman, Karpinski, and Shah 2017) are two high-level programming languages that are used in particular for statistics and numerical analysis. Connecting Julia and R is particularly interesting because the two languages can complement each other. While the history of R dates back to 1976 (Becker and Chambers 1984), the first Julia release was in 2013 (Shah 2013). As can be expected with its long history, R offers a much larger number of packages for statistical methods than Julia. Yet, Julia also has packages that offer features that are not available in R. For example, training neural differential equations (Chen, Rubanova, Bettencourt, and Duvenaud 2018), which will be shown in an example later, is not directly possible in R at the moment. Another example where R users can benefit from developments in Julia is the Julia package **DifferentialEquations**, which also is available via the wrapper package **diffeqr** in R (Rackauckas and Nie 2017). Julia was created with a strong emphasis on computational speed, as the authors were not satisfied with the performance of existing scientific computing languages (Bezanson *et al.* 2017). This makes Julia particularly interesting for statisticians who have to deal with computationally intensive tasks. Notable Julia packages that exploit the performance advantages of Julia are the optimization package **JuMP** (Dunning, Huchette, and Lubin 2017), and the **MixedModels** package (Bates 2020), which we will use for a small introductory example below.

There are already two packages, **JuliaCall** (Li 2019) and **XRJulia** (Chambers 2016, Chapter 15), that aim at integrating Julia in R. This demonstrates the interest of the community in making functionality from Julia available in R. The requirements of convenient interactive use, e.g., of Julia deep learning packages, led us to develop the new package **JuliaConnectoR** for this purpose.

While Julia can get close to the performance of C, it offers the convenience of a high-level scripting language, which allows for fast code development. This makes connecting with Julia an alternative to using C extensions in R. Considering the similarities between R and Julia, this can further aid in making computationally demanding algorithms available in R. Also, there are already many bridges to different languages available (Dahl 2020; Urbanek 2009; Allaire, Ushey, Tang, and Eddelbuettel 2017), making R very suitable as a glue language. This ability to interface with other software is even central to the design of R (Chambers 2016).

While R offers a wide range of statistical approaches, access to deep learning techniques is mostly provided only by wrappers around packages from other languages. Examples for this are the R packages **keras** (Allaire and Chollet 2019), and **rTorch** (Reyes 2019), which wrap the Python libraries **Keras** (Chollet *et al.* 2015) and **PyTorch** (Paszke, Gross, Massa, Lerer, Bradbury, Chanan, Killeen, Lin, Gimelshein, Antiga, Desmaison, Kopf, Yang, DeVito, Raison, Tejani, Chilamkurthy, Steiner, Fang, Bai, and Chintala 2019), respectively. These two packages employ Python via the language bridge provided by the R package **reticulate** (Allaire *et al.* 2017). Julia also offers an innovative approach for deep learning via the native Julia package **Flux** (Innes 2018). **Flux** is based on differentiable programming, a technique for interpreting programs as functions which can be differentiated. The gradients of these functions are calculated via automatic differentiation, which can happen at execution time or even at compile time. Thus, it becomes less necessary for deep learning developers to adapt to a particular programming style that is enforced by a specific framework, e.g., computational graphs in **TensorFlow** (Abadi, Isard, and Murray 2017). Instead, the optimization can be performed automatically on typical code. Julia is particularly suited to support this (Innes, Karpinski, Shah, Barber, Stenetorp, Besard, Bradbury, Churavy, Danisch, Edelman, Malmaud, Revels, and Yuret 2018). We designed our package having deep learning with Julia in mind. The goal is to be able to interact with Julia code in a natural way in R.

Julia and R both target users in the fields of statistics and machine learning. This is mirrored by the fact that both languages share more traits with each other than with languages such as C or Python, which have not been designed primarily for this user group. In R, a functionally oriented programming style is more common than object-oriented programming (OOP), even if there are several different OOP approaches available in R (Wickham 2019). Similar to R, Julia is also not focused on OOP. Instead, Julia relies on "multiple dispatch" (Bezanson *et al.* 2017) as the central paradigm, where appropriate methods are selected based on the types of all arguments at runtime. The generic function OOP found in both R and Julia is different from an encapsulated object-oriented style (Wickham 2019), e.g., employed in Python or Java. The interface of the **JuliaConnectoR** reflects such commonalities between Julia and R, leveraging the parallels between the two languages as much as possible.

In the following, we provide a first introductory example that shows how we can use the **JuliaConnectoR** for conducting statistical analyses with R and Julia, before discussing package features in more detail, and subsequently providing a deep learning example.

## 2. Introductory example

For a first look at a typical workflow with the **JuliaConnectoR**, we consider analyses with mixed models. In R, mixed models can be fitted, e.g., with the **lme4** package (Bates, Mächler, Bolker, and Walker 2015). One of the authors of the mixed models package **lme4** in R also develops the Julia package **MixedModels**. An advantage of the Julia package is that it can often fit models much faster. To demonstrate this, let us examine the data set `InstEval` from the **lme4** package, which contains data about evaluations of university lectures, collected over several years. The numeric evaluation score is contained in the variable `y`. The variable `d` contains an identification number of individual professors or lecturers. Likewise, the variables `s` and `dept` contain identifiers for students and departments, respectively. The variable `service` denotes whether the lecture was a service lecture, i.e., if the lecturer held the lecture for a different department than his own. An exemplary goal could be to find the lecturer with the highest score, adjusting for the effects of the department. For this, we can compare the random effects of the lecturers in a mixed model. To do this in R, we can use the following code:

```
R> library("lme4")
R> load("InstEval.rda")
R> formula1 <- y ~ 1 + service*dept + (1|s) + (1|d)
R> fm1 <- lmer(formula1, InstEval, REML = TRUE)
R> which.max(ranef(fm1)$d$`(Intercept)`)

[1] 665
```

For brevity, the code snippet omits the preparation of the data set and simply loads the table from the file `InstEval.rda`. The pre-processing steps are documented in the supplementary material. After the data has been loaded, the model formula is defined and the linear mixed model is fitted via the function `lmer` from the **lme4** package. The `ranef` function can finally be used to extract the random effects estimators from the model to find their largest value.

To fit the same model by calling out to the Julia package **MixedModels**, we can use the following code:

```
R> library("JuliaConnectoR")
R> MM <- juliaImport("MixedModels")
R> RData <- juliaImport("RData")
R> InstEval <- RData$load("InstEval.rda")[["InstEval"]]
R> formula2 <- juliaEval(
+    "MixedModels.@formula(y ~ 1 + service*dept + (1|s) + (1|d))")
R> fm2 <- MM$fit(MM$LinearMixedModel, formula2, InstEval, REML = TRUE)
R> which.max(juliaCall("MixedModels.ranef", fm2)[[2]])
```

```
[1] 665
```

At first we also need to load the data set. R datasets can be loaded with the **RData** Julia package (Bates, White, and Stukalov 2020b). To connect to Julia, we use our **JuliaConnectoR** package. The **JuliaConnectoR** does not know about **RData** or **MixedModels**. But the `juliaImport` function can scan any given package for everything that is a function or can act as a function. All these objects are bundled in an environment and returned by `juliaImport`. Calling or referencing a function or other object from a Julia package is then possible via accessing the returned environment. With this we can access the `load` function from the **RData** package to get the data set in Julia. Defining the formula in Julia is possible with the `juliaEval` function, which evaluates arbitrary Julia code and returns the result to R. To minimize the communication overhead, complex objects are by default passed as references to Julia objects. Fitting the model is also possible in a straightforward way using the imported functions and objects. After precompilation, fitting the model in Julia is much faster than in R (e.g., 1.5 versus 10 seconds on a typical PC). In addition to using `juliaImport` and the function references, it is possible to call functions by their name via the function `juliaCall`. In our example, the analyses using Julia reproduce the results from the **lme4** package.

## 3. Features

In the following, we describe the most important features of the **JuliaConnectoR**. In this process, we also highlight parallels and differences between Julia and R. We also compare the **JuliaConnectoR** package (version 0.6) to the packages **JuliaCall** (version 0.17.1) and **XRJulia** (version 0.9.0) with respect to each of the described features.

A short overview of the comparison can be seen in Table 1.

| *Feature* | **JuliaConnectoR** | **JuliaCall** | **XRJulia** | See |
|---|---|---|---|---|
| Communication | TCP/binary | C-interface | TCP/JSON | 3.1 |
| Automatic importing of packages | Yes | No | No | 3.2 |
| Specification for type translation | Yes | No | No | 3.3 |
| Reversible translation from Julia to R | Yes | No | No | 3.4 |
| Callbacks | Yes | Yes | No | 3.5 |
| Let-syntax | Yes | No | No | 3.6 |
| Show standard (error) output | Yes | No | Yes | 3.7 |
| Interruptible | Yes | No | Yes | 3.8 |
| Missing values | Yes | Yes | No | 3.9 |
| R data frames to Julia Tables | Yes | Yes | No | 3.10 |

Table 1: Comparison of features between the **JuliaConnectoR** (version 0.6), **JuliaCall** (version 0.17.1) and **XRJulia** (version 0.9.0).

### 3.1. Communication protocol

The **JuliaConnectoR** starts a Julia process in the background and communicates with it via a custom binary protocol that is based on the transmission control protocol (TCP) (Postel 1981). Interprocess-communication via TCP is an established technique, which is used by Julia itself and also by R packages integrating external machine learning systems such as **h2o** (Aiello, Eckstrand, Fu, Landry, and Aboyoun 2018) or **sparklyr** (Luraschi, Kuo, Ushey, Allaire, Macedo, Falaki, Wang, Zhang, Li, Hajnala, Szymkiewicz, Davis, RStudio, Inc., and The Apache Software Foundation 2020). The communication protocol of the **JuliaConnectoR** uses messages, which may contain arbitrarily complex nested objects. The format is inspired by BSON (MongoDB, Inc. 2009), a format that is an alternative binary format to JavaScript Object Notation (JSON) (Bray 2017). Like BSON, the **JuliaConnectoR** serialization format uses the binary form of the objects directly, exploiting the commonalities of the binary representations between Julia and R. On the R side, the `writeBin` and `readBin` functions can directly write and read whole R arrays. In Julia the `write` and `read` methods for binary IO (input/output) connections can be used. This avoids transformations which are necessary for using text-based exchange formats like JSON, where numbers have to be converted to strings containing decimal representations. The protocol also allows streaming the messages, which means that messages do not have to be read completely, but a simultaneous writing and reading/parsing is possible. This allows for fast and efficient communication.

**XRJulia** lets R and Julia communicate via JSON messages, which can encapsulate vectors, matrices and also more complex nested structures. Due to the conversion issues of JSON mentioned above, **XRJulia** deviates from the strategy of serializing everything in JSON format by writing large vectors and matrics in intermediate files in a binary format (see `largeVectors` documentation item in the manual of **XRJulia**). Yet, writing files to the hard drive for communicating is still slower than sending files via TCP. Using files is also an obstacle for taking advantage of the potential of TCP-based communication, which is the ability to potentially run Julia as a server and R as a client on different machines or containers.

**JuliaCall** connects Julia and R using the Julia package **RCall** (Bates, Lai, Byrne, and contributors 2020a). **RCall** integrates R and Julia using C interfaces. From a technical perspective, this is a tighter integration. Earlier versions were rather unstable and not always compatible with different Julia versions. This was one of the reasons why we started developing our own interface. Coupling Julia and R via their C interfaces makes communication faster, but the looser coupling via TCP also has benefits: First, it makes developing and maintaining the package much easier. In particular, the quick release cycles of Julia exacerbate problems of maintenance. With a coupling on a higher level of the interfaces, the compatibility has a higher chance of surviving an update. Additionally to supporting different versions, it is furthermore possible to support a wider range of configurations with this. For example, a particular configuration requirement of **RCall** is that R has to be compiled with the `-enable-R-shlib` option to build R as a dynamic library [1]. But such a setup is not wanted in all cases, as it can reduce the performance of R by 10-20 % (see R Core Team (2020b), Appendix B.1).

### 3.2. Automatic importing of whole packages and modules

One main feature of the **JuliaConnectoR** is that it can import whole packages from Julia conveniently in one command. The function `juliaImport` scans specified packages for functions and types and creates corresponding R functions, which are returned bundled in an enviroment. Types are also imported as functions, as they can be used as constructors or functors. The information that these functions are constructors for types is attached as attribute in

---

[1] https://github.com/JuliaInterop/RCall.jl/blob/v0.13.4/docs/src/installation.md

R, such that these type-constructor functions can be passed as type arguments to other Julia functions.

Translating Julia code into R code thus becomes straightforward. Consider the following snippet of Julia code, which loads the Julia package manager **Pkg**, installs the **Flux** Julia package in version 0.11, and defines a small neural network using **Flux**:

```
julia> import Pkg
julia> Pkg.add(name = "Flux", version = "0.11")
julia> import Flux
julia> model = Flux.Chain(Flux.Dense(4, 4, Flux.relu), Flux.Dense(4, 1))
```

This can be translated into the following R code:

```
R> Pkg <- juliaImport("Pkg")
R> Pkg$add(name = "Flux", version = "0.11")
R> Flux <- juliaImport("Flux")
R> model <- Flux$Chain(Flux$Dense(4L, 4L, Flux$relu), Flux$Dense(4L, 1L))
```

In addition to importing installed packages, it is also possible to load plain modules from source code. This is particularly useful when one wants to interactively develop Julia code in parallel to R code. A good workflow for developing Julia code is to put functions into modules. For loading a module in the current session, the corresponding (main) file can be executed via the Julia function `include`. This can be repeated in a Julia session multiple times, with the module being replaced completely every time. Thereby, the workspace can be kept clean. With the **JuliaConnectoR**, this workflow is also possible when working in an R session. If one follows this strategy and has, e.g., the Julia functions in a module `MyModule` defined in the file `mymodule.jl`, importing the module can be done via using its relative module path like in Julia:

```
R> juliaCall("include", "/path/to/mymodule.jl")
R> MyModule <- juliaImport(".MyModule")
```

If the module is the last thing that is defined in the file, it is returned as result of evaluating the file, and importing can be done in one line:

```
R> MyModule <- juliaImport(juliaCall("include", "/path/to/mymodule.jl"))
```

**JuliaCall** can import functions of packages via the function `julia_pkg_import`. This function does not scan packages, but the names of the functions need to be specified. **XRJulia** has functions `juliaImport` and `juliaUsing`, but those behave differently and do not return or assign functions. The mechanism of connecting to Julia functionality also is more complex in **XRJulia**.

### 3.3. Translation from R to Julia

Julia is more sensitive to types than R. In contrast to R, Julia allows to specify types of arguments in functions. Functions that use this Julia feature then only accept arguments of specific types. On the one hand, being specific about types has advantages: It allows the Julia compiler to create efficient code. It also makes it possible to dispatch on the type. This means that one function can have multiple methods depending on the types of its arguments, which may be optimized for handling different types in the most efficient way. Julia then infers automatically which is the most specific method to pick. On the other hand, many R users may not be used to thinking about types, as most R functions handle types in a very relaxed way. So, it is worthwhile to take a look at how the **JuliaConnectoR** translates types between Julia and R.

The basic R types and their counterparts in Julia are shown in Table 2. Vectors containing only one element are translated to the type shown in the table. R arrays with more than one element and/or having a dimension specified via the `dim` attribute are translated to Julia `Array`s of the corresponding type and dimension. For example, the R integer vector `c(1L, 2L)` will be of type `Array{Int,1}` in Julia. A double matrix such as `matrix(c(1, 2, 3, 4), nrow = 2)` will be of type `Array{Float64,2}`. The translation is performed implicitly when passing values to Julia. If, however, the same large R matrix needs to be accessed in multiple calls to Julia, the function `juliaPut` may be used to translate/copy an R vector or matrix to Julia only once. The resulting proxy object can then be referenced in multiple calls to Julia without requiring further copying of data.

| R | → | Julia |
|---|---|---|
| integer | → | Int |
| double | → | Float64 |
| logical | → | Bool |
| character | → | String |
| complex | → | Complex{Float64} |
| raw | → | UInt8 |

Table 2: Basic R types and their corresponding Julia types

More complex R data structures can also be translated: R `list`s are translated to Julia objects of type `Array{T, 1}`, where `T` is the most specific Julia type of the translated elements contained in the list. This works with arbitrarily nested lists.

Data frames are handled in a special way, see below in section 3.10.

Even if this translation may be intuitive to users familiar with types in R and Julia, the clear specification of the translated types is a feature of the **JuliaConnectoR** that helps with common type issues, since it makes the type of arguments predictable. More details regarding this can be found in the package documentation.

**JuliaCall** and **XRJulia** currently lack such a clear specification. From experiments it seems that **JuliaCall** and **XRJulia** use the same mapping of the types as specified in Table 2, although this is not documented. **XRJulia** 0.9.0 failed to translate complex values. (E.g., `juliaCall("typeof", 1i)` returned an error).

### 3.4. Translation from Julia to R

The type system of Julia is richer than that of R. The **JuliaConnectoR** follows the principle that data structures translated from Julia should be reconstructable with their original type if needed. This eases the integration of Julia code that relies on specific types. For example, the documentation of **Flux** recommends to use single-precision floating point values (Julia type `Float32`) for performance reasons. If one were to translate a matrix with elements of type `Float32` to an R `double` array, add no type information, and then use it again with **Flux**, the inferred type for the call would be `Float64`: The code would lose its speed advantage unless the type was managed explicitly. For handling this, the **JuliaConnectoR** adds the type information as an attribute to translated objects. To ensure a minimal distraction on the command line output, the type is only added if it is needed. The translations resulting from following this principle are shown in Table 3.

| Julia | $\rightarrow$ | R |
| --- | --- | --- |
| `Float64` | $\rightarrow$ | `double` |
| `Float16`, `Float32`, `UInt32` | $\rightarrow$ | `double` with type attribute |
| `Int64` that fits in 32 bits | $\rightarrow$ | `integer` |
| `Int64` not fitting in 32 bits | $\rightarrow$ | `double` with type attribute |
| `Int8`, `Int16`, `UInt16`, `Int32`, `Char` | $\rightarrow$ | `integer` with type attribute |
| `UInt8` | $\rightarrow$ | `raw` |
| `UInt64`, `Int128`, `UInt128`, `Ptr` | $\rightarrow$ | `raw` with type attribute |
| `Complex{Float64}` | $\rightarrow$ | `complex` |
| `Complex{Int{X}}` with $X \leq 64$ | $\rightarrow$ | `complex` with type attribute |
| `Complex{Float{X}}` with $X \leq 32$ | $\rightarrow$ | `complex` with type attribute |
| `String` | $\rightarrow$ | `character` |

Table 3: Julia types that are directly translated to R by the **JuliaConnectoR**

Julia functions are translated to R functions that call the respective Julia functions. With this, an anonymous function can be defined in Julia and assigned in R:

```
R> f2 <- juliaEval("x -> x .^ 2")
```

The same can be done with a named function:

```
R> f2 <- juliaEval('function f2(x)
+                     x .^ 2
+                  end')
```

If a named function exists already, it can be imported directly via `juliaFun`:

```
R> f2 <- juliaFun("f2")
```

In any case, the resulting R function can be used like any R function:

```
R> f2(2)
```

```
[1] 4
```

Julia provides Unicode support and strings are encoded in UTF-8 in Julia by default. It is furthermore possible to use non-ASCII characters in variable names in Julia and there are people who make use of this feature. For example, the Julia package **Flux** uses Greek letters and mathematical symbols ($\sigma$, $\nabla$) in function names. In R, the native encoding depends on the locale, and Windows does not yet support UTF-8 locales (Kalibera 2020). The **JuliaConnectoR** translates strings to UTF-8 encoded strings in R. These are available regardless of the native encoding but there may be issues arising when processing strings further as implicit conversions of strings in R may lead to unexpected results (Kalibera 2020). For the compatibility across different platforms, it is therefore advisable to refrain from using other characters than those expressible in US-ASCII encoding as far as it is possible. If functions from packages that make use of non-ASCII variable names are needed, the **JuliaConnectoR** facilitates writing cross-platform code by providing alternative names for functions when importing them via `juliaImport`. These alternative names make use of LaTeX-like abbreviations that are defined in Julia for enabling tab completion sequences to type special characters on the command line. For example, the Julia function `Flux.log`$\sigma$ is available as `Flux$log`$\sigma$ on UTF-8 locales and as `` Flux$`log<sigma>` `` on all locales and platforms if the package has been imported, like shown in the example in section 3.2.

Objects that are more complex than `Array`s of above types are returned to R in the form of proxy objects. These proxy objects can be examined and used in place of the original objects in subsequent calls to Julia. Unnecessary copying of complex objects is avoided by communicating only references to the original Julia objects to R for constructing the proxy objects. Examples for objects that are not converted by default are Julia `struct` types or arrays of arrays. This behavior allows to get an easy access to simple objects, which are straightforward to use in R. It also allows to handle more complex objects in a performant and safe way, including objects having references to external resources, such as pointers to memory or file handles.

A full translation of complex objects to Julia is possible via the `juliaGet` function. Julia objects that do not contain circular references or external pointers can be reconstructed from their translations to R. Such objects can thus also be serialized together with the R session.

This is implemented based on the translation of complex Julia `Array`s and `struct`s to R `list`s. In the case of `struct`s, the names of the list elements correspond to the field names of the struct. Retaining the type information as an attribute in R, the original objects can be assembled again.

Consider the following Julia code in a file `MyLibrary.jl`, defining a struct `Book`.

```julia
module MyLibrary
  export Book, cite

  struct Book
    author::String
    title::String
    year::Int
  end

  function cite(book::Book)
    "$(book.author): $(book.title) ($(book.year))"
  end
end
```

Such a `struct` can be fully translated to an R `list`, which can again be translated back to a Julia object when passed to a Julia function:

```
R> MyLibrary <- juliaImport(juliaCall("include", "/path/to/MyLibrary.jl"))
R> book <- juliaGet(MyLibrary$Book("Shakespeare", "Romeo and Julia", 1597L))
R> book

$author
[1] "Shakespeare"

$title
[1] "Romeo and Julia"

$year
[1] 1597

attr(,"JLTYPE")
[1] "Main.MyLibrary.Book"

R> MyLibrary$cite(book)

[1] "Shakespeare: Romeo and Julia (1597)"
```

For comparison, **JuliaCall** also creates proxy objects for more complex objects, e.g., for an array of arrays such as `[[1;2], [3;4]]`. But there is no possibility for an automatic translation of such a structure to a native R data structure.

**XRJulia** also has a function `juliaGet`, which can translate more complex structures to R. However, it does not annotate the translation with the original types, so an exact reconstruction is generally not possible.

### 3.5. Callbacks

R functions are translated to Julia functions that call the original R functions. This way, they can be passed to Julia functions as arguments. It is possible to nest callbacks, e.g., invoking a Julia function that calls an R function as a callback that again may call a Julia function, and so on. This feature makes the **JuliaConnectoR** a truly functional interface. This kind of communication becomes possible by the custom TCP-based protocol, which allows bidirectional communication. A simple example using the Julia function `map` (which is analogous to the R function `Map`) demonstrates this:

```
R> juliaCall("map", function(x) {x+1}, c(1,2,3))

[1] 2 3 4
```

Such callback functions are useful, e.g., for monitoring the training progress when training a neural network. We demonstrate this below with the neural ordinary differential equation example (see section 4).

**JuliaCall** can also use R functions in place of Julia, functions. For example, `julia_call("map", function(x) x+1, c(1,2,3))` returns the same result. **XRJulia** does not allow to pass R functions as arguments. It also does not translate Julia functions to R functions.

### 3.6. `let` syntax

The usage of the keyword `let` in functionally oriented programming languages is inspired by mathematical language. In Julia, `let` allows to create a new scope, and declare variables in this scope. The value of the expression is the value of the final expression in the block. From this perspective, a `let` block is equivalent to defining an anonymous function and evaluating it only once. Using a `let` block in Julia, declaring (intermediate) global variables can be avoided, which allows for a clean programming style.

The function `juliaLet` of the **JuliaConnectoR** allows to create such a `let` block in a simple way. R variables can be passed as arguments. They are then inserted for the corresponding variables in the expression, which is given as a string. With this, one can evaluate complex Julia expressions and insert R variables in place of Julia variables.

The primary usage scenario for `juliaLet` is to build Julia objects using Julia syntax from R data without having to define intermediate variables at the global scope. The following code demonstrates this by creating a dictionary object, using distinct Julia syntax, in a straightforward way:

```
R> juliaLet('Dict("x" => x, "y" => y)', x = c(1,2), y = c(2,3))

<Julia object of type Dict{String,Array{Float64,1}}>
Dict("x" => [1.0, 2.0],"y" => [2.0, 3.0])
```

**JuliaCall** and **XRJulia** do not have an equivalent function.

### 3.7. Output redirection

The standard output and standard error output from Julia are redirected and displayed in the R console. This is particularly useful for interactive work because warnings or output are needed to detect errors. The implementation of this is challenging because catching the output needs to be handled asynchronously from the function call itself. We implement this with Julia `Task`s that collect the output. These tasks are synchronized in Julia, so that R can receive the messages synchronously.

**JuliaCall** does not have this feature. Recent versions of **XRJulia** are able to display the output.

### 3.8. Interrupting

In interactive programming of machine learning algorithms, it is important to be able to interrupt long-running commands. The **JuliaConnectoR** catches interrupts that are signaled via `Ctrl + C` key or `esc` in RStudio and terminates the running Julia process.

To test this feature, we can run an infinite loop and try to interrupt it.

```
R> juliaEval('while true; end')
```

It is currently not possible to interrupt a command in **JuliaCall**. For example, RStudio as a development environment needed to be shut down forcefully after executing `julia_eval('while true; end')`. Interrupting commands in **XRJulia** is possible.

### 3.9. Missing values

Julia has a concept of missing values with a three-valued logic like R. The difference is that missing values are of the distinct type `Missing`, which has the single value `missing`. This means an array that may contain missing values has a different type than an array without missing values. For example, `[1.0; 2.0; 3.0]` is of type `Array{Float64,1}` and `[1.0; missing; 3.0]` is of type `Array{Union{Missing, Float64},1}`. In R, on the other hand, `c(1.0, 2.0, 3.0)` and `c(1.0, NA, 3.0)` have the same type. Both behaviors are integrated by the **JuliaConnectoR**. Missing values in R (value `NA`) are translated to `missing` values in Julia. R arrays with missing values are converted to Julia arrays of type `Array{Union{Missing, T}}`, where `T` stands for the translated type in Table 2. `NA` and `NaN` values are distinguished. A `double` value of `NaN` in R is translated to a `NaN` value of type `Float64` in Julia and vice versa.

```
R> juliaCall("+", c(1, NA, NaN), c(1, 2, 3))
```

```
[1]   2  NA NaN
```

```
R> juliaCall("sqrt", NA)
```

```
[1] NA
```

**JuliaCall** also supports missing values. The corresponding command `julia_call("+", c(1, NA), c(1, 2))` yields the same result as the code above. But a fatal error occurs when calling `julia_call("sqrt", NA)` (on version 0.17.1), which terminates the R session.

**XRJulia** does not handle `NA`s properly. The first command from the code snippet above returns a proxy object pointing to an object of type `Array{Float64, 1}` which cannot be retrieved via `juliaGet`. The second command prints a warning message and returns `NULL`.

### 3.10. Data frame support

Data of tabular shape is very common for statistics. Therefore, this is a very important type of data structure for languages like R and Julia, which both focus on making statistical and numerical computing easily accessible.

Unsurprisingly, the implementation of such tabular data has some parallels in R and Julia. Base R provides *data frames*, but there are some packages providing alternative versions of such data structures, namely **tibble** and **data.table**. The implementations of tables provided by these packages extend the basic data frame interface, such that these tables can be used like a `data.frame` from base R.

Similar to R, there are different takes on data frames in Julia, provided by different packages, most prominently **DataFrames** (Julia Data collaborators 2020) and **JuliaDB** (Julia Computing, Inc. 2020). There is also a unifying interface for the data structures defined in these packages, which is defined in the **Tables** Julia package.

These parallels are used in the **JuliaConnectoR** to make it simple to exchange tabular data. By enabling translations between the interfaces `data.frame` and **Tables**, it becomes possible to translate other kinds of tabular data structures extending `data.frame`s, in particular `tibble`s and `data.table`s, as well. For implementing the **Tables** interface, we use a custom type, which wraps the columns in a minimal way.

As information about the internal structure of the original Julia objects gets lost with the translation to R, it is generally not possible to fully restore an object implementing the **Tables** interface from its `data.frame` translation. For this reason, the resulting tabular data structures are given back to R as proxy objects. Performance is another reason: By using proxy objects, it is possible to interactively create large subsets of Julia Tables without having to worry about heavy traffic between Julia and R because only references need to be transferred instead of the complete data. The translation of data structures from Julia to R can be requested via a call to `as.data.frame`.

**JuliaCall** translates data frames to `DataFrame` objects from the **DataFrames** package. This is also compatible with the **Tables** interface. We decided against **DataFrames** since it is more heavy-weight and also has not reached version 1.0 yet.

**XRJulia** translates data frames to a proxy object encapsulating a `Dict{String, Any}`. This has the disadvantage that the column order is lost and it is also not compatible to the **Tables** interface.

## 4. An example using neural differential equations

As a further illustration, we use the **JuliaConnectoR** to reconstruct a deep learning approach proposed by Chen *et al.* (2018) in their work on neural ordinary differential equations, which won a best paper award at NeurIPS 2018 (https://nips.cc/Conferences/2018/Awards). Specifically, the authors present a new family of deep learning models by combining neural networks with ordinary differential equations.

Many types of neural networks such as residual networks (He, Zhang, Ren, and Sun 2016) or recurrent neural networks (RNNs) (Rumelhart, Hinton, and Williams 1986) are characterized by applying a finite sequence of discrete transformations to a hidden state $h_t$:

$$h_{t+1} = h_t + f(h_t, \theta_t), \quad t \in \{0, \ldots, T\}$$

The central idea in the work of Chen *et al.* (2018) is to generalize this discretized transformation to a continuous dynamic of the hidden units in the form of an ordinary differential equation (ODE):

$$\frac{dh(t)}{dt} = f(h(t), \theta_t), \quad t \in [0, T] \quad (1)$$

The function $f(h(t), \theta_t)$ that parameterizes the derivative of the hidden state is given by a neural network with the parameters $\theta_t$. To obtain the value of the hidden state at some time $T$, instead of applying all transformations for $t = 0, 1, \ldots, T$, the initial value problem defined by equation (1) and a starting point $h(0)$ can be solved at $T$ using a black-box differential equation solver.

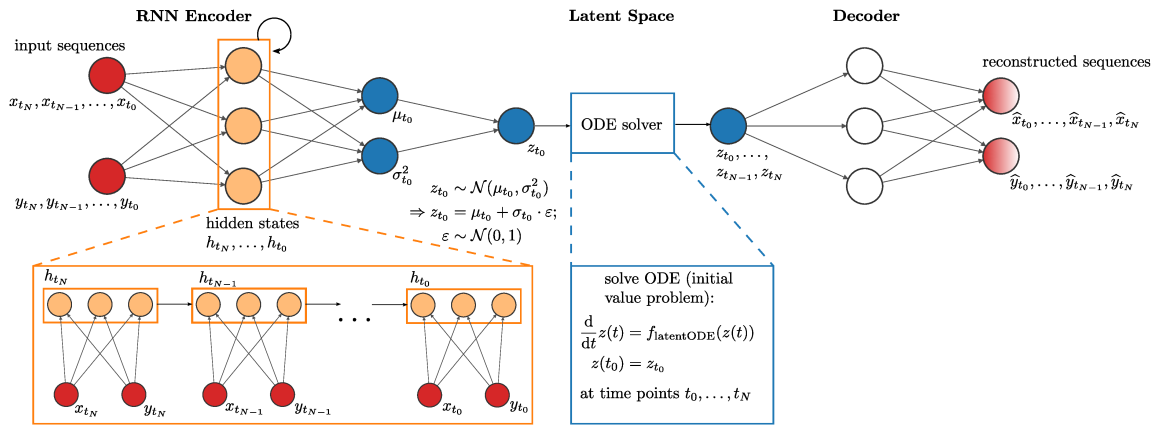

Figure 1: Overview of the architecture for modeling the trajectories of a series $(x_{t_0}, y_{t_0}), \ldots, (x_{t_N}, y_{t_N})$ of two-dimensional points in time. The model architecture is based on a variational autoencoder, consisting of an encoder network, a decoder network, and a latent space with reduced dimensionality in between.

In the paper, the authors propose a memory-efficient way of solving the ODE by using the adjoint sensitivity method, a technique for performing reverse-mode differentiation (aka backpropagation) through the ODE solver with constant memory cost with respect to the network depth. As a result, this allows to build continuous-depth residual networks and continuous-time latent variable models that can be trained efficiently. The concrete model architecture, shown in Figure 1, is based on a variational autoencoder (VAE), a generative deep learning model first presented by Kingma and Welling (2014). It consists of two distinctly parameterized but jointly optimized neural networks: The encoder maps the input data $x$ to a lower-dimensional latent representation $z$, while the decoder performs the reverse transformation from the latent space back to data space. The encoder and decoder parameterize the conditional distributions $q_\phi(z|x)$ and $p_\theta(x|z)$, respectively. This formulation as probability distributions allows to generate synthetic data after training by drawing samples from the learned distributions. The VAE training objective is to recover the central factors of variation underlying the data in the lower-dimensional latent space, thus obtaining a compressed representation. Based on the framework of variational inference (Blei, Kucukelbir, and McAuliffe 2017), a loss function for the VAE can be derived, which is given by the negative of the *evidence lower bound* (ELBO):

$$\mathcal{L}_{\text{VAE}}(x, \phi, \theta) = -\text{ELBO}(x, \phi, \theta) = D_{\text{KL}}(q_\phi(z|x) \| p(z)) - \mathbb{E}_{q_\phi(z|x)}[\log(p_\theta(x|z))] \qquad (2)$$

Intuitively, the second term of the loss function can be thought of as a reconstruction error. The first term, in which $D_{KL}$ denotes the Kullback-Leibler divergence, enforces the consistency between the prior $p(z)$ (assumed to follow a standard normal distribution) and the posterior distribution of $z$. It can be shown (Blei *et al.* 2017) that

$$\log p(x) \geq \text{ELBO}(x, \phi, \theta).$$

This means minimizing the VAE loss amounts to maximizing a lower bound on the true data likelihood.

The model from Chen *et al.* (2018) is trained on time series data. These time series are represented as latent trajectories in the model, where each trajectory is obtained by solving an ODE system in the latent space of the VAE model. Our specific dataset comprises time series of two-dimensional points $(x_{t_0}, y_{t_0}), \ldots, (x_{t_N}, y_{t_N})$ that are drawn from a spiral trajectory. More specifically, there are two underlying spiral trajectories, one clockwise and one counter-clockwise, each including 100 two-dimensional points. These points can be thought of as a time series of coordinates of a point on the two-dimensional plane that travels inwards along a spiral-shaped trajectory. From these underlying trajectories, we generate 100 training observations by randomly sampling a starting point somewhere on the trajectory and adding Gaussian noise with mean 0 and standard deviation 0.1 to the subsequent 20 points, corresponding to the next 20 time points of the trajectory. Each observation thus consists of a time series of 20 points moving inwards along the trajectory of one of the two spirals (clockwise or counter-clockwise). The complete code for creating the data set is available as supplementary material. The underlying ground-truth spiral trajectories and the samples of two time-series observations from the training set are shown in Figure 2.

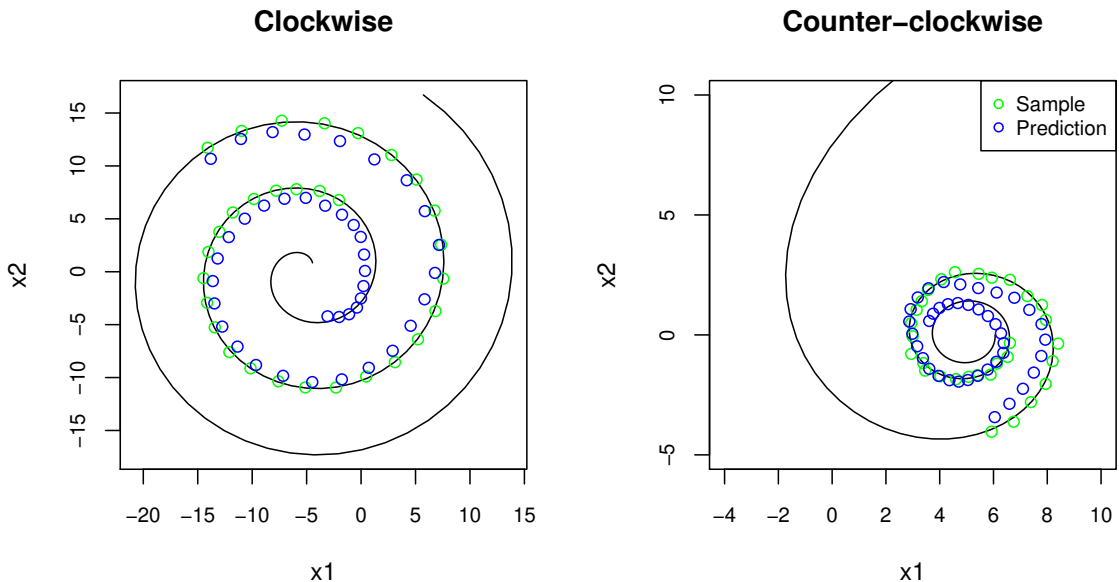

Figure 2: Learning of samples from spiral trajectories with the latent time series VAE. The points on the spiral are moving inwards over time. Half of the samples are drawn from the clockwise spiral and half from the counter-clockwise one. The trajectories can be predicted by solving the differential equation in the VAE latent space and can be extrapolated by solving it at other time points than the ones in the training data.

We want to demonstrate how to implement such a complex deep learning model in R. This also more generally shows how the **JuliaConnectoR** can be used to bring capabilities for deep learning and related novel techniques from Julia to R. To implement the model for capturing these patterns in Julia, we use the machine learning packages **Flux** (Innes 2018) and **DiffEqFlux** (Rackauckas, Innes, Ma, Bettencourt, White, and Dixit 2019). **DiffEqFlux** integrates the deep learning models from **Flux** with differential equations and realizes efficient backpropagation through arbitrary ODE solvers.

The Julia code for the example can be found in the file `SpiralExample.jl`, which defines the Julia module `SpiralExample`. The R code that uses the Julia functions of the module is contained in the file `SpiralExample.R`. We first consider the implementation of the model architecture in Julia. The model is composed of different parts, which are collected in a Julia data structure of type `LatentTimeSeriesVAE`:

```
struct LatentTimeSeriesVAE
  rnn
  latentODEfunc
  decoder
end
```

For constructing the model, which is depicted in Figure 1, we can specify the dimensions following the architecture from Chen *et al.* (2018). Specifically this corresponds to the following parameters (with the actual numbers given in parentheses):

- the number of dimensions in the latent space, i.e., the length of $\mu$ and $\sigma$ (4),
- the number of observed dimensions (2),
- the number of hidden nodes for the RNN encoder (25),
- the number of hidden dimensions for the neural network defining the latent ODE function (20), and
- the number of hidden nodes for the decoder network (20).

These parameters affect the ability of the model to learn the structure in a data set. Therefore, it is useful to be able to set the parameters easily for tuning the model. For this purpose, we define a constructor method, which accepts these parameters and returns an initialized network:

```
function LatentTimeSeriesVAE(; latent_dim, obs_dim,
                               rnn_nhidden, f_nhidden, dec_nhidden)

  rnn = Chain(RNN(obs_dim, rnn_nhidden), Dense(rnn_nhidden, latent_dim*2))

  latentODEfunc = Chain(Dense(latent_dim, f_nhidden, Flux.elu),
                        Dense(f_nhidden, f_nhidden, Flux.elu),
                        Dense(f_nhidden, latent_dim))

  decoder = Chain(Dense(latent_dim, dec_nhidden, Flux.relu),
                  Dense(dec_nhidden, obs_dim))
  LatentTimeSeriesVAE(rnn, latentODEfunc, decoder)
end
```

The training of the model is performed in the Julia function `SpiralExample.train!`. Besides taking care of some monitoring and data preparation, the actual training is performed with the **Flux** function `train!`, which requires a loss function that is to be minimized. In our case, the loss function is the negative of the ELBO (plus a regularization term concerning the weights of the RNN). The ELBO is calculated in the `elbo` function in the `SpiralExample` module:

```
function elbo(model::LatentTimeSeriesVAE, x::SpiralSample, t)
  empmu, emplogsd = latent_mu_logsd(model, x)
  Flux.reset!(model.rnn)
  z0 = latentz0(empmu, emplogsd)
  pred_z = n_ode(model, z0, t)
  sumlogp_x_z = sum([logp_x_z(x[i],
    model.decoder(pred_z[:,i])) for i in 1:size(pred_z,2)])
  sumlogp_x_z - kl_q_p(empmu, emplogsd)
end
```

To keep the code clean, the different calculations in the `elbo` function are outsourced to helper functions in the `SpiralExample` module: In `latent_mu_logsd`, $\mu$ and $\log \sigma$ are calculated from the input to the encoder network `x`. The latent variable $z$ is then sampled from the normal distribution $\mathcal{N}(\mu, \sigma)$. From the ODE solver, we get the predicated values for $z$ via `n_ode`. With this, the terms of the ELBO can finally be calculated via the helper functions `kl_q_p` and `logp_x_z`, which are used for determining the values of the two terms in formula (2).

In the next step, we show how to use this model in R. This also illustrates some more general patterns which typically occur in code that performs deep learning.

Although the default installation of Julia offers a broad range of features in the standard library, advanced functionality, such as algorithms for deep learning, needs to be added via external packages. The Julia installation is discovered by the **JuliaConnectoR** from the system environment variable `JULIA_BINDIR`, which may define the directory that contains the Julia executable, or it is found by looking in the executable path for a "julia" executable. A proper setup of the basic Julia installation can be ensured via the function `juliaSetupOk`. Extra packages can be added via the Julia package manager, which can be accessed via the **JuliaConnectoR**. In case of this example, we additionally need the Julia packages **Flux**, **DiffEqFlux**, and their dependencies. To ensure optimal reproducibility of the experiment, we use a Julia "project". This is simply a directory containing a `Projects.toml` file and a `Manifest.toml` file. These files specify the exact versions of all packages used. They are created by Julia to record the state of all package operations that have been executed while using a project. With the Julia function `Pkg.activate`, we tell Julia to use the project. With `Pkg.instantiate`, we can install all project dependencies with one call.

```
R> Pkg <- juliaImport("Pkg")
R> Pkg$activate(".")
R> Pkg$instantiate()
```

In the `SpiralExample` Julia module, a type `LatentTimeSeriesVAE` is defined, which combines several neural networks models from the package **Flux** in a single object. After importing the module (see also section 3.2), we can call the constructor of this type to initialize the model, specifying the parameters of the architecture. The resulting object is by default translated to an R object that serves as proxy for the corresponding Julia object. The `model` proxy object holds only a reference to the Julia object, and no data are copied to R, which makes this approach also feasible and effective for large networks.

```
R> juliaCall("include", normalizePath("SpiralExample.jl"))
R> SpiralExample <- juliaImport(".SpiralExample")
R> model <- SpiralExample$LatentTimeSeriesVAE(latent_dim = 4L,
+    obs_dim = 2L, rnn_nhidden = 25L, f_nhidden = 20L, dec_nhidden = 20L)
```

It is very common in deep learning that many parameters are used. To avoid confusion, it is advised to use named arguments, as shown above. In contrast to R, named arguments and positional arguments are strictly separated in Julia. For named arguments, Julia requires the names and does not infer their values by their position.

In the next step, the model can be trained on the data:

```
R> epochs <- 20
R> plotValVsEpoch <- function(epoch, val) {
+    if (epoch == 1) {
+      ymax <- max(val)
+      plot(x = 1, y = val,
+        xlim = c(0, epochs), ylim = c(0, ymax*1.1),
+        xlab = "Epoch", ylab = "Value")
+    } else {
+      points(x = epoch, y = val)
+    }
+  }
R> spiraldata <- SpiralExample$spiral_samples(nspiral = 100L,
+    ntotal = 150L, nsample = 30L, start = 0, stop = 6*pi, a = 0, b = 1)
R> SpiralExample$`train!`(model, spiraldata$samp_trajs, spiraldata$samp_ts,
+    epochs = epochs, learningrate = 0.01, monitoring = plotValVsEpoch)
```

The training function receives the `model`, the $x$ values in `spiraldata$samp_trajs` and the time values in `spiraldata$samp_ts`. Optional named arguments can be specified here as well. The example also demonstrates the use of a callback function, which is here specified as `monitoring` argument. The `plotValVsEpoch` defined above is used to plot the value of the loss function during the training. It is called after each training `epoch`. (An "epoch" is deep learning jargon for the update of the model parameters based on the loss function using all samples once.) A plot of the loss function (see Figure 3) allows to evaluate the training progress at one glance. Displaying output directly during the training is especially useful if the training takes longer.

After the model has been trained, we can evaluate the model performance. In our case, we take a look at the model predictions for the training observations. By parameterizing the latent space dynamics as a time-continuous ODE solution, we can inter- and extrapolate the time series by solving the ODE at other time points than the ones observed in the training data and decoding them to data space (see Figure 2). The prediction can be done, e.g., with the following code:

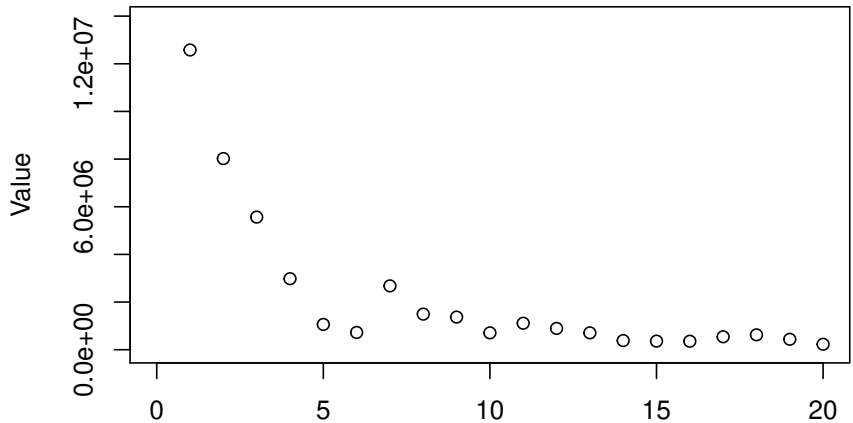

Figure 3: The output of the function `plotValVsEpoch`, which is used to plot the value of the loss over the number of epochs. In this case, the loss gets suddenly worse in one step but decreases to a better optimum later in the training.

```
R> predlength <- length(spiraldata$samp_ts) + 10
R> SpiralExample$predictspiral(
+    model, sample, spiraldata$orig_ts[1:predlength])
```

Here, the `sample` contains the $x$ values, and `spiraldata$orig_ts` contains all possible time values.

## 5. Summary and outlook

We have introduced the **JuliaConnectoR** package for connecting R and Julia in a reliable and convenient way. The example with neural differential equations shows how the **JuliaConnectoR** can help to enable more flexible ways of deep learning in R. It also demonstrates some best practices for employing Julia, such as using Julia modules.

The comparison of current features in Table 1 provides an overview over the different language bridges between Julia and R. It can be seen that the **JuliaConnectoR** is a new solution for connecting Julia and R that offers many features not available in other packages. The usage of the most important functions of the package has been exemplified in the small code snippets for illustrating the features. An overview of the functions that have been presented here can be seen in Table 4.

There is also some potential that is not yet fully leveraged: A next step for developing the **JuliaConnectoR** further could be to harness the fact that the communication via TCP could in principle be used to run Julia sessions remotely from the computer running the R session. This might be useful when users want to use a convenient UI for programming on their machine, and at the same time, they would like to utilize resources on remote computing servers. As detailed in section 3.1, the **JuliaConnectoR** is the only one of the three packages connecting Julia and R whose design allows it to be used in such a way. It is technically not complicated to run a Julia server with the Julia part of the **JuliaConnectoR** on a remote server in the network, and connect to it from a different computer. Yet, additional security measures need to be implemented for putting this in practice since the execution of arbitrary code can be triggered via such a connection.

| Function name | Short description | Usage see |
|---|---|---|
| `juliaImport` | Load a Julia package in Julia via `import` and return its functions and data types as an environment, such that the functions can be called directly in R | 3.2 |
| `juliaFun` | Create an R function that wraps a Julia function | 3.4 |
| `juliaCall` | Call any Julia function by name. (Not needed for functions created via `juliaImport` or `juliaFun`.) | 3.2, 3.9 |
| `juliaEval` | Evaluate a simple Julia expression (and return the result) | 3.4, 3.8 |
| `juliaLet` | Evaluate Julia expressions with R variables in place of Julia variables employing a `let` block (and return the result) | 3.6 |
| `juliaGet` | Fully translate a Julia object to an R object | 3.4 |

Table 4: Overview of most important exported functions

The goal of the current version is to connect R and Julia in an intuitive way. The best example for this is the automatic importing of Julia packages. Also, the features for interactive use, such as the redirection of the standard (error) output and the possibility to interrupting running commands, make it easier to develop extensions for R in Julia.

Additionally, the **JuliaConnectoR** comes with a design that aims to avoid state in Julia that is not visible in R: Julia functions can be translated to R functions and all variables returned from Julia are translated into R variables. It is not necessary or encouraged by the package to use global variables. In places where Julia and R do not align so easily, this is aided by the introduction of the function `juliaLet` to handle more complex Julia expressions. By that, the design of the **JuliaConnectoR** allows for a clean style of programming and minimizes the feeling of "remote-controlling" Julia.

# Acknowledgements

This work has been supported by the Federal Ministry of Education and Research (BMBF) in Germany in the MIRACUM project (FKZ 01ZZ1801B).

# References

Abadi M, Isard M, Murray DG (2017). "A Computational Model for **TensorFlow**: An Introduction." *Proceedings of the 1st ACM SIGPLAN International Workshop on Machine Learning and Programming Languages*, pp. 1–7. doi:10.1145/3088525.3088527.

Aiello S, Eckstrand E, Fu A, Landry M, Aboyoun P (2018). "Machine Learning with R and **H2O**." *H2O booklet*, **550**. URL https://www.h2o.ai/resources/.

Allaire JJ, Chollet F (2019). **keras**: *R Interface to 'Keras'*. R package version 2.2.5.0, URL https://CRAN.R-project.org/package=keras.

Allaire JJ, Ushey K, Tang Y, Eddelbuettel D (2017). **reticulate**: R *Interface to* Python. URL https://github.com/rstudio/reticulate.

Bates D (2020). "Julia Package **MixedModels** (GitHub Repository)." doi:10.5281/zenodo.3727845. URL https://github.com/JuliaStats/MixedModels.jl.

Bates D, Lai R, Byrne S, contributors (2020a). "Julia Package **RCall** (GitHub Repository)." URL https://github.com/JuliaInterop/RCall.jl.

Bates D, Mächler M, Bolker B, Walker S (2015). "Fitting Linear Mixed-Effects Models Using **lme4**." *Journal of Statistical Software*, **67**(1), 1–48. ISSN 1548-7660. doi:10.18637/jss.v067.i01. Number: 1.

Bates D, White JM, Stukalov A (2020b). "Julia Package **RData** (GitHub Repository)." URL https://github.com/JuliaData/RData.jl.

Becker RA, Chambers JM (1984). S: *an Interactive Environment for Data Analysis and Graphics*. Belmont, Calif. : Wadsworth Advanced Book Program.

Bezanson J, Edelman A, Karpinski S, Shah V (2017). "Julia: A Fresh Approach to Numerical Computing." *SIAM Review*, **59**(1), 65–98. ISSN 0036-1445. doi:10.1137/141000671.

Blei DM, Kucukelbir A, McAuliffe JD (2017). "Variational Inference: A Review for Statisticians." *Journal of the American Statistical Association*, **112**(518), 859–877. doi:10.1080/01621459.2017.1285773.

Bray T (2017). *The* JavaScript *Object Notation (JSON) Data Interchange Format*. Library Catalog: tools.ietf.org, URL https://tools.ietf.org/html/rfc8259.

Chambers JM (2016). *Extending* R. CRC Press, Boca Raton, Florida.

Chen TQ, Rubanova Y, Bettencourt J, Duvenaud D (2018). "Neural Ordinary Differential Equations." In *Advances in Neural Information Processing Systems*.

Chollet F, *et al.* (2015). "**Keras**: the Python Deep Learning API." https://keras.io.

Dahl D (2020). "Integration of R and Scala Using **rscala**." *Journal of Statistical Software*, **92**(4), 1–18. ISSN 1548-7660. doi:10.18637/jss.v092.i04.

Dunning I, Huchette J, Lubin M (2017). "**JuMP**: A Modeling Language for Mathematical Optimization." *SIAM Review*, **59**(2), 295–320. doi:10.1137/15M1020575.

He K, Zhang X, Ren S, Sun J (2016). "Deep Residual Learning for Image Recognition." In *Proceedings of the IEEE Conference on Computer Vision and Pattern Recognition (CVPR)*.

Innes M (2018). "**Flux**: Elegant Machine Learning with Julia." *Journal of Open Source Software*, **3**(25), 602. doi:10.21105/joss.00602.

Innes M, Karpinski S, Shah V, Barber D, Stenetorp P, Besard T, Bradbury J, Churavy V, Danisch S, Edelman A, Malmaud J, Revels J, Yuret D (2018). "On Machine Learning and Programming Languages." In *SysML Conference 2018*. URL https://mlsys.org/Conferences/doc/2018/37.pdf.

Julia Computing, Inc (2020). "Julia Package **JuliaDB** (GitHub Repository)." URL https://github.com/JuliaComputing/JuliaDB.jl.

Julia Data collaborators (2020). "Julia Package **DataFrames** (GitHub Repository)." URL https://github.com/JuliaData/DataFrames.jl.

Kalibera T (2020). "UTF-8 Support on Windows - The R Blog." URL https://developer.r-project.org/Blog/public/2020/05/02/utf-8-support-on-windows/.

Kingma DP, Welling M (2014). "Auto-Encoding Variational Bayes." In Y Bengio, Y LeCun (eds.), *2nd International Conference on Learning Representations (ICLR), Conference Track Proceedings*.

Li C (2019). "**JuliaCall**: an R Package for Seamless Integration between R and Julia." *Journal of Open Source Software*, **4**(35), 1284. doi:10.21105/joss.01284.

Luraschi J, Kuo K, Ushey K, Allaire J, Macedo S, Falaki H, Wang L, Zhang A, Li Y, Hajnala J, Szymkiewicz M, Davis W, RStudio, Inc, The Apache Software Foundation (2020). **sparklyr**: *R Interface for* **Apache Spark**. R package version 1.5.2, URL https://CRAN.R-project.org/package=sparklyr.

MongoDB, Inc (2009). "BSON (Binary JSON) Serialization." URL http://bsonspec.org/.

Paszke A, Gross S, Massa F, Lerer A, Bradbury J, Chanan G, Killeen T, Lin Z, Gimelshein N, Antiga L, Desmaison A, Kopf A, Yang E, DeVito Z, Raison M, Tejani A, Chilamkurthy S, Steiner B, Fang L, Bai J, Chintala S (2019). "**PyTorch**: An Imperative Style, High-Performance Deep Learning Library." In H Wallach, H Larochelle, A Beygelzimer, F d'Alché-Buc, E Fox, R Garnett (eds.), *Advances in Neural Information Processing Systems 32*, pp. 8024–8035. Curran Associates, Inc. URL http://papers.neurips.cc/paper/9015-pytorch-an-imperative-style-high-performance-deep-learning-library.pdf.

Postel J (1981). *Transmission Control Protocol*. Library Catalog: tools.ietf.org, URL https://tools.ietf.org/html/rfc793.

R Core Team (2020a). *R: A Language and Environment for Statistical Computing*. R Foundation for Statistical Computing, Vienna, Austria. URL https://www.R-project.org/.

R Core Team (2020b). *R Installation and Administration*. R version 4.0.3, URL https://cran.r-project.org/doc/manuals/r-release/R-admin.pdf.

Rackauckas C, Innes M, Ma Y, Bettencourt J, White L, Dixit V (2019). "**DiffEqFlux.jl** - A Julia Library for Neural Differential Equations." 1902.02376, URL https://arxiv.org/abs/1902.02376.

Rackauckas C, Nie Q (2017). "**DifferentialEquations.jl** – a Performant and Feature-Rich Ecosystem for Solving Differential Equations in Julia." *Journal of Open Research Software*, **5**(1).

Reyes AR (2019). **rTorch**: *R Bindings to 'PyTorch'*. R package version 0.0.3, URL https://CRAN.R-project.org/package=rTorch.

Rumelhart D, Hinton G, Williams R (1986). "Learning Representations by Back-Propagating Errors." *Nature*, **323**, 533–536. `doi:10.1038/323533a0`.

Shah VB (2013). "Julia 0.1 Release on GitHub." URL `https://github.com/JuliaLang/julia/releases/tag/v0.1`.

Urbanek S (2009). **rJava**: *Low-Level R to Java Interface*. R package version 0.8-1, URL `https://CRAN.R-project.org/package=rJava`.

Wickham H (2019). *Advanced R*. 2 edition. CRC Press, Boca Raton, Florida. ISBN 978-1-4665-8696-3.